\documentclass[pdflatex,sn-mathphys-num]{sn-jnl}

\usepackage{graphicx}%
\usepackage{multirow}%
\usepackage{amsmath,amssymb,amsfonts}%
\usepackage{amsthm}%
\usepackage{mathrsfs}%
\usepackage[title]{appendix}%
\usepackage{xcolor}%
\usepackage{textcomp}%
\usepackage{manyfoot}%
\usepackage{booktabs}%
\usepackage{algorithm}%
\usepackage{algorithmicx}%
\usepackage{algpseudocode}%
\usepackage{listings}%
\usepackage{hyperref}
\usepackage{placeins}
\usepackage{rotating}
\usepackage{float}

\theoremstyle{thmstyleone}%
\theoremstyle{thmstyletwo}%

\theoremstyle{thmstylethree}%

\begin{document}

\title[Article Title]{Compact Neural Network Algorithm for Electrocardiogram Classification}


\author[1]{\fnm{Chrstian M.} \sur{Frausto-Avila}}
\equalcont{These authors contributed equally to this work.}

\author[3]{\fnm{Juan P.} \sur{Manriquez-Amavizca}}
\equalcont{These authors contributed equally to this work.}

\author[1]{\fnm{Ana K. S.} \sur{Rocha Robledo}}

\author[1]{\fnm{Mario A.} \sur{Quiroz-Juarez}}\email{maqj@fata.unam.mx}

\author[2]{\fnm{Alfred B.} \sur{U'Ren}}\email{alfred.uren@correo.nucleares.unam.mx}

\affil[1]{\orgdiv{Centro de F\'{i}sica Aplicada y Tecnolog\'{i}a Avanzada}, \orgname{Universidad Nacional Aut\'onoma de M\'exico}, \orgaddress{\street{Boulevard Juriquilla 3001}, \city{Quer\'{e}taro}, \postcode{76230}, \state{Quer\'{e}taro}, \country{ M\'exico}}}

\affil[2]{\orgdiv{Instituto de Ciencias Nucleares}, \orgname{Universidad Nacional Aut\'onoma de M\'exico}, \orgaddress{\street{04510}, \city{Quer\'{e}taro}, \postcode{70543}, \state{Quer\'{e}taro}, \country{ M\'exico}}}

\affil[3]{\orgdiv{Instituto Tecnol\'ogico y de Estudios Superiores de Monterrey}, \orgname{Universidad Nacional Aut\'onoma de M\'exico}, \orgaddress{\street{Epigmenio Gonz\'alez 500, Fracc, San Pablo}, \city{Quer\'{e}taro}, \postcode{76130}, \state{Quer\'{e}taro}, \country{ M\'exico}}}


\abstract{In this paper, we present a powerful, compact electrocardiogram (ECG) classification algorithm for cardiac arrhythmia diagnosis that addresses the current reliance on deep learning and convolutional neural networks (CNNs) in ECG analysis. This work aims to reduce the demand for deep learning, which often requires extensive computational resources and large labeled datasets. Our approach introduces an artificial neural network (ANN) with a simple architecture combined with advanced feature engineering techniques. A key contribution of this work is the incorporation of 17 engineered features that enable the extraction of critical patterns from raw ECG signals. By integrating mathematical transformations, signal processing methods, and data extraction algorithms, our model captures the morphological and physiological characteristics of ECG signals with high efficiency, without requiring deep learning. Our method demonstrates a similar performance to other state-of-the-art models in classifying 4 types of arrhythmias, including atrial fibrillation, sinus tachycardia, sinus bradycardia, and ventricular flutter. Our algorithm achieved an accuracy of 97.36\% on the MIT-BIH and St. Petersburg INCART arrhythmia databases. Our approach offers a practical and feasible solution for real-time diagnosis of cardiac disorders in medical applications, particularly in resource-limited environments.}




\maketitle

\section{Introduction}\label{sec1}

Arrhythmic diseases are defined as irregularities in the normal functioning of the electrical impulses that are responsible for regulating the heartbeat. Arrhythmias can manifest in various forms and trigger multiple health issues. Some arrhythmias are benign, while others may represent life-threatening risks. The World Health Organization (WHO) indicates that cardiovascular diseases remain the leading cause of death worldwide. They account for 31\% of all global deaths \cite{singh2023ecg,mendis2015organizational}. Consequently, the study and development of methodologies for the detection of arrhythmias through technological devices are crucial for maintaining public health and reducing the risk of fatal cardiac disorders. The electrocardiogram (ECG) is a widely employed tool that plays an essential role in this important task. First introduced in 1902 by Willem Einthoven \cite{alghatrif2012brief}, ECG reflects the macroscopic electrical activity of the heart, generating a visual record of this activity as a function of time. This electrical activity governs the contractions and relaxations of the heart muscle that occur with each heartbeat. Due to its non-invasive nature, ECG recordings are currently the most widely used method for evaluating a patient's cardiovascular health and are considered a universal standard in clinical practice and cardiological studies. Approximately, 300 million ECGs are recorded every year \cite{sattar2023electrocardiogram}. Importantly, the relevance of ECGs has reached the academic world. Significant research efforts have been dedicated to modeling and understanding ECG signals, in order to study and comprehend the macroscopic electrical activity of the heart \citep{ quiroz2019periodically, gois2009analysis, quiroz2019generation, mcsharry2003dynamical, quiroz2018cardiac}. 

\begin{figure}[t!] 
    \centering
    \includegraphics[width=8cm]{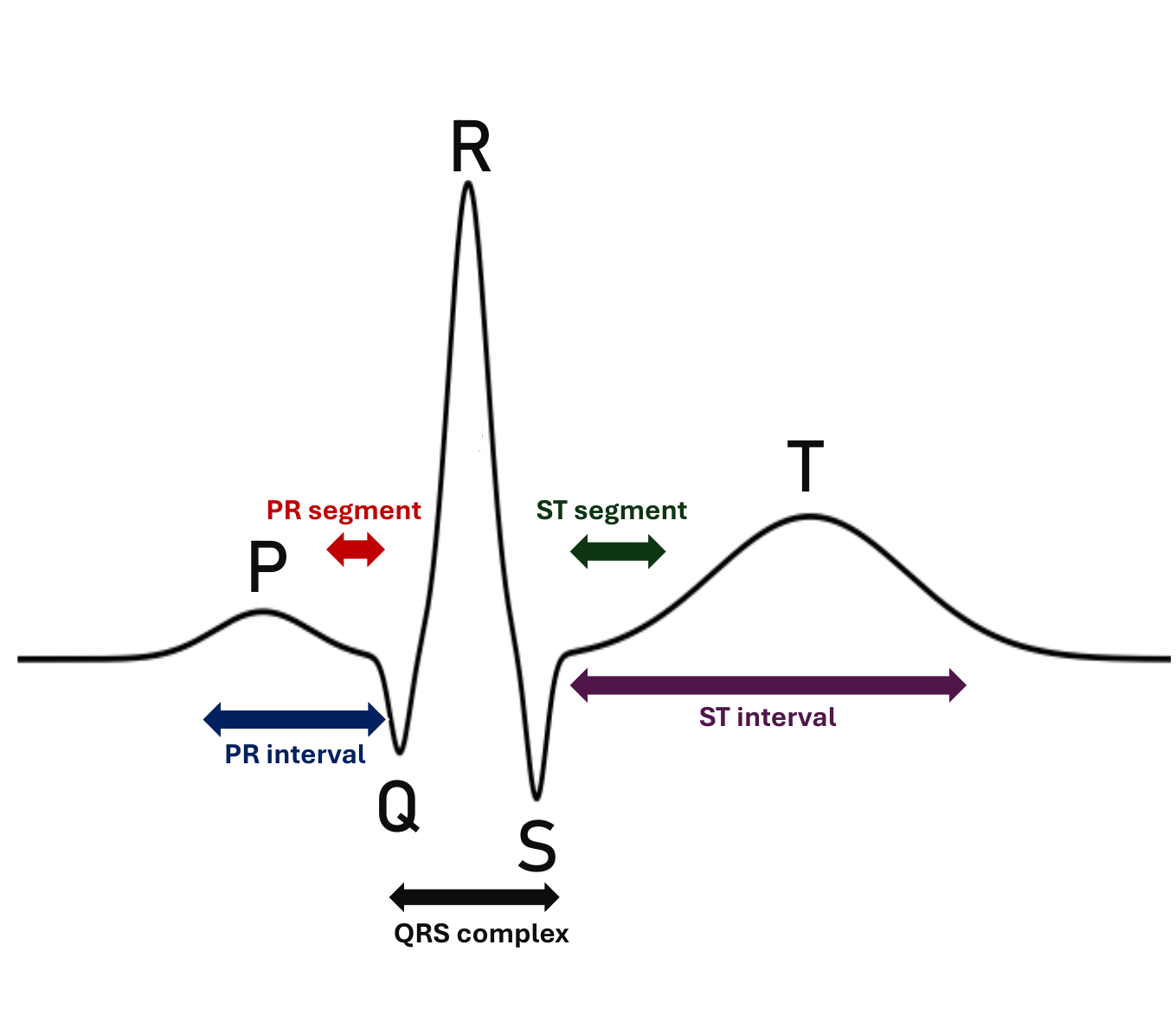}  
    \caption{Typical PQRST pattern present in a healthy body.}
    \label{fig:ecgsignal}
\end{figure}

Under normal conditions, the ECG signal follows a characteristic PQRST waveform, which represents the sequential depolarization and repolarization events regulating heart function. Figure \ref{fig:ecgsignal} illustrates the peaks and troughs of the ECG waveform, labeled P, Q, R, S, and T. This pattern is standard in healthy individuals, and deviations from it may indicate cardiac conditions such as arrhythmias. ECG is a reliable tool for detecting abnormalities in the PQRST pattern and diagnosing various cardiac disorders, including tachycardia (heart rate >100 BPM), bradycardia (<60 BPM), hypertension, hypotension, and myocardial infarction \citep{khan2023ecg, yildirim2018arrhythmia}. It is also effective in identifying other arrhythmias such as atrial and ventricular fibrillation, congenital heart defects, heart block, coronary artery disease, pericarditis, cardiomyopathy, electrolyte imbalances, and rheumatic heart disease. However, ECG-based diagnosis has limitations. Interpretation often relies on the clinician’s expertise, and factors like noise or similarities between arrhythmic disorders can complicate the analysis. Moreover, the same arrhythmia can manifest with different symptoms and rhythm abnormalities, posing a challenge even for experienced cardiologists. Given these challenges, developing novel techniques to enhance and simplify cardiac data analysis is essential.

Notably, machine learning (ML) has emerged as a powerful tool for identifying arrhythmic disorders from clinical ECG signals. Its ability to recognize intricate patterns in complex data has led to significant advancements across various scientific disciplines \cite{ML_Intr_1, villegas2022identification, lollie2022high, salazar2024improving}. In ECG classification tasks, convolutional neural networks (CNNs) have been employed to enhance classification accuracy. For instance, Ref. \cite{8356203} proposed a CNN that automates feature extraction, further streamlining the analysis process. Similarly, Jun et al. \cite{jun2018ecg} developed a two-dimensional CNN that achieved an impressive accuracy of 99.05\%. Wang et al. \cite{wang2021automatic} introduced another CNN architecture that incorporates the continuous wavelet transform (CWT) as an input feature, achieving a classification accuracy of 98.74\%. Note that these models eliminate the need for manual feature engineering and are inherently more robust to noise. However, deep learning approaches require large datasets and significant computational resources for training, posing challenges for real-world implementation.

Alternatively, advanced feature engineering techniques have been proposed to extract relevant information and intricate patterns from raw ECG signals. For instance, Sadhukhan et al.  \cite{sadhukhan2018automated} applied a discrete Fourier transform (DFT) to the ECG signal as an engineered feature for an ML model, achieving a performance of 95.6\%. However, this approach was limited to the detection of myocardial infarction. Likewise, Zeng et al. \cite{zeng2024detection} implemented the Shannon energy envelope as an extracted feature, reaching an accuracy of 99.21\%, though it was also restricted to detecting a single cardiac disorder. Other methods for ECG signal classification involve approaches based on support vector machines (SVM), decision tree algorithms, and K-Nearest Neighbors (KNN) classifiers. Rabee and Barhumi \cite{SVM_ECG} developed an SVM-based approach for the recognition of 14 distinct heartbeats from the MIT-BIH arrhythmia database. The authors applied a wavelet transform to the signal for feature extraction. Their classifier achieved a 99.2\% accuracy. Kumari et al. \cite{DT_ECG} engineered a decision tree to classify six types of heartbeats, including normal rhythm and five arrhythmic types. The authors obtained an accuracy of 94.5\%, after extracting 17 morphological features from the signals. 

Mohebbanaaz and Padma \cite{KNN_ECG} developed a KNN-based approach and achieved a classification of 99.03\% using the MIT-BIH database with 6 types of ECG beats. Normal sinus rhythm (NSR), left bundle branch block (LBBB), right bundle branch block (RBBB), premature ventricular contraction (PVC), atrial premature beat (APB), and paced beat (PAB). However this approach required a boosting algorithm for hyperparameter optimization that increased the complexity of the whole methodology, without which the accuracy drop to 94.5\%. Nevertheless, these methods demonstrate the importance of feature engineering for ECG signal classification, which plays a crucial role in early detection and diagnosis of cardiovascular diseases, requiring efficient and accurate machine learning models. This approach has allowed us to engineer a more robust classification algorithm. 

\begin{figure*}[t!]
    \centering
    \includegraphics[width=\linewidth]{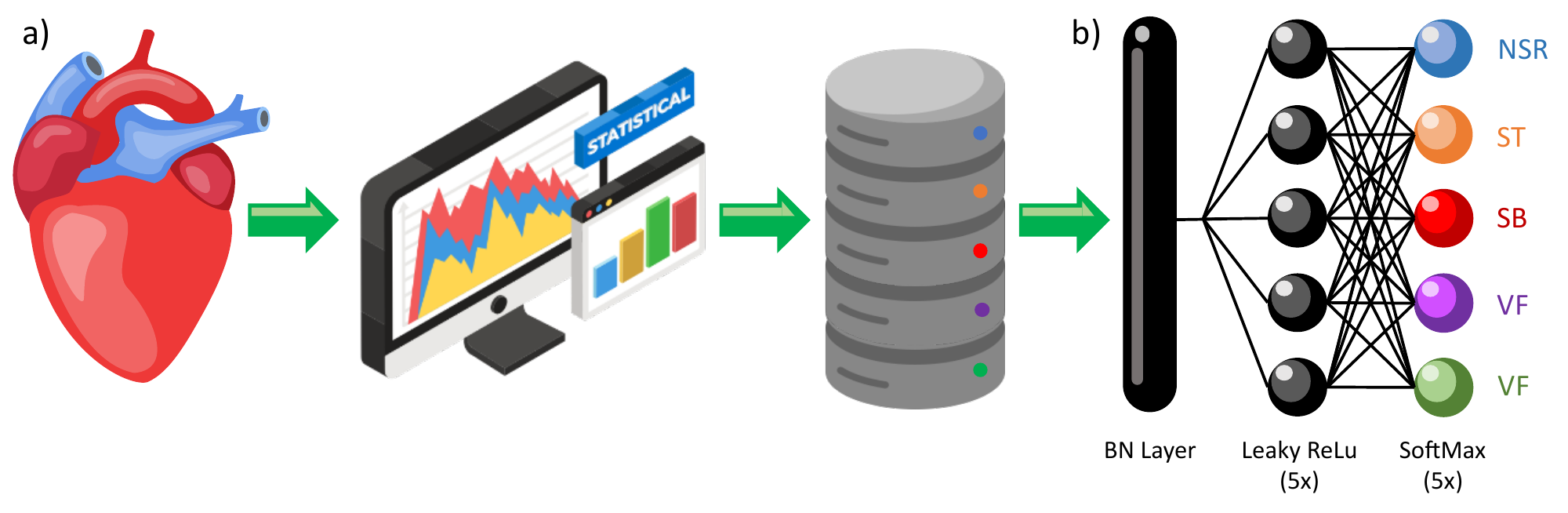} 
    \caption{General scheme of our approach. a) Real recordings obtained from patients at both the Beth Israel Deaconess Medical Center and St. Petersburg Institute of Cardiological Technics,  Feature engineering of the ECG recording and the database creation from Normal Sinus Rhythm, SB to Sinus Bradycardia, ST to Sinus Tachycardia, VF to Ventricular Flutter, and AF to Atrial Fibrillation and b) Neural Network architecture.}
    \label{fig:Scheme}
\end{figure*}

Our study has been significantly motivated by recent advancements in deep learning. These studies have demonstrated large potential in medical diagnostics using machine learning algorithms and data extraction techniques. One particularly influential approach is triplet representation learning, as explored by \cite{ren2024exploring}. This technology is designed to improve feature engineering and data extraction processes through the implementation of the SimTrip representation learning model. This methodology has been successfully implemented for medical image classification and it demonstrated its effectiveness in learning discriminative features for high-dimensional and complex medical data. Inspired by this scientific advancement, we have integrated advanced feature extraction techniques in our ECG classification algorithm study to improve robustness, generalizability and a lower computational complexity. By carefully implementing data extraction principles principles, we aim to improve the model’s ability to distinguish subtle variations in ECG signals by capturing important morphological and statistical features. This approach has ultimately allowed us to engineer a more robust classification algorithm.  Additionally, our work has been influenced by \cite{ren2022hybrid}. this remarkable study highlights the how generative adversarial networks (GANs) are able to generate synthetic lung cancer images and extract key features for their classification. The study incorporates CNNs to extract intricate morphological information from complex medical data. CNNs have  extensively demonstrated how deep learning models can be employed for feature extraction in biomedical engineering applications and diagnostic tools. As a result, we focus on developing a streamlined approach for feature extraction using mathematical transformations and data extraction techniques like Principal Component Analysis (PCA). Rather than relying solely on deep learning networks,  our method integrates feature engineering with a compact neural network architectures to maximize the efficiency of ECG classification while maintaining model simplicity.  Furthermore, our work has been motivated by recent advancements in large foundation models for medical imaging, as explored by \cite{foundation_cancer_segmentation}. This study has demonstrated the capacity of efficient neural networks in processing complex medical data. These models exhibit the potential of feature-efficient architectures in handling diverse and high-dimensional medical data with complex and intricate patterns hidden within them. 

In this study, we propose a compact neural network–based classification framework for the detection of four arrhythmic disorders using real clinical electrocardiogram (ECG) records. The low computational complexity of the model makes it particularly well suited for integration into portable medical devices and mobile health applications. The framework incorporates advanced feature engineering strategies to substantially reduce computational overhead, including statistical descriptors that capture ECG morphological and waveform characteristics, as well as signal transformations such as the Hilbert and Fourier transforms to extract complex, high-order patterns. The resulting low-dimensional neural network attains an overall classification accuracy of 97.36\%, achieving a favorable balance between computational efficiency and diagnostic performance. These findings underscore the relevance of the proposed approach to ECG signal analysis and its potential for enabling real-time, resource-constrained mobile healthcare systems.

\section{Methodology}

The goal of our work is to classify arrhythmic diseases from real ECG signal data through the implementation of a minimally complex neural network algorithm. To achieve this, we first built a database of ECG recordings from real clinical records to train our model. Next, we applied advanced and novel feature engineering techniques to extract key characteristics and intricate patterns from the ECG signals as shown in Fig. \ref{fig:Scheme}. The proposed features served as inputs to the ML model.  Finally, the algorithm undergoes hyperparameter optimization to maximize its predictive performance. Fig. \ref{fig:Scheme} shows a general scheme of our approach for the ECG classification task.

\subsection{Database}

Our initial efforts focused on collecting data from real ECG clinical records. This work focuses on the classification of four types of arrhythmic disorders: Sinus Tachycardia (ST), Sinus Bradycardia (SB), Atrial Fibrillation (AF) and Ventricular Flutter (VF). The dataset is composed of ECG recordings from the open MIT-BIH Arrhythmia Database, which is available at the PhysioNet platform and contains data from 47 subjects \cite{8356203}. These recordings were initially obtained from patients at the Beth Israel Deaconess Medical Center, located at Boston. The recordings were part of a series of cardiology experiments conducted between 1975 and 1979 \cite{moody2001impact}. Moreover, the database includes 23 randomly selected recordings from a total of 4,000 24-hour ambulatory ECG recordings collected at the same institution. These additional recordings contain rare and clinically significant arrhythmic diseases for academic and research purposes. For this work, we used all available recordings.  All samples have a duration of 30 minutes. They were originally recorded at 360 Hz, with 11-bit resolution over a 10 mV dynamic range. In order to increase the size of our dataset and diversify the sources of the ECG signals used to train the algorithm, we incorporated ECG recordings from the open St. Petersburg INCART 12-lead Arrhythmia Database, which is also available at PhysioNet. The source is composed of 75 recordings from 32 patients collected at the St. Petersburg Institute of Cardiac Technics \cite{goldberger2000physiobank}. These recordings were originally sampled at 257 Hz. Each recording contains 12 different signals that correspond to a different electrode placement.  Note that in this study, they used the standard Lead I, which consists of placing the positive electrode on the left arm and the negative electrode on the right arm. 

\begin{figure*}[t]
    \centering
    \includegraphics[width=\linewidth]{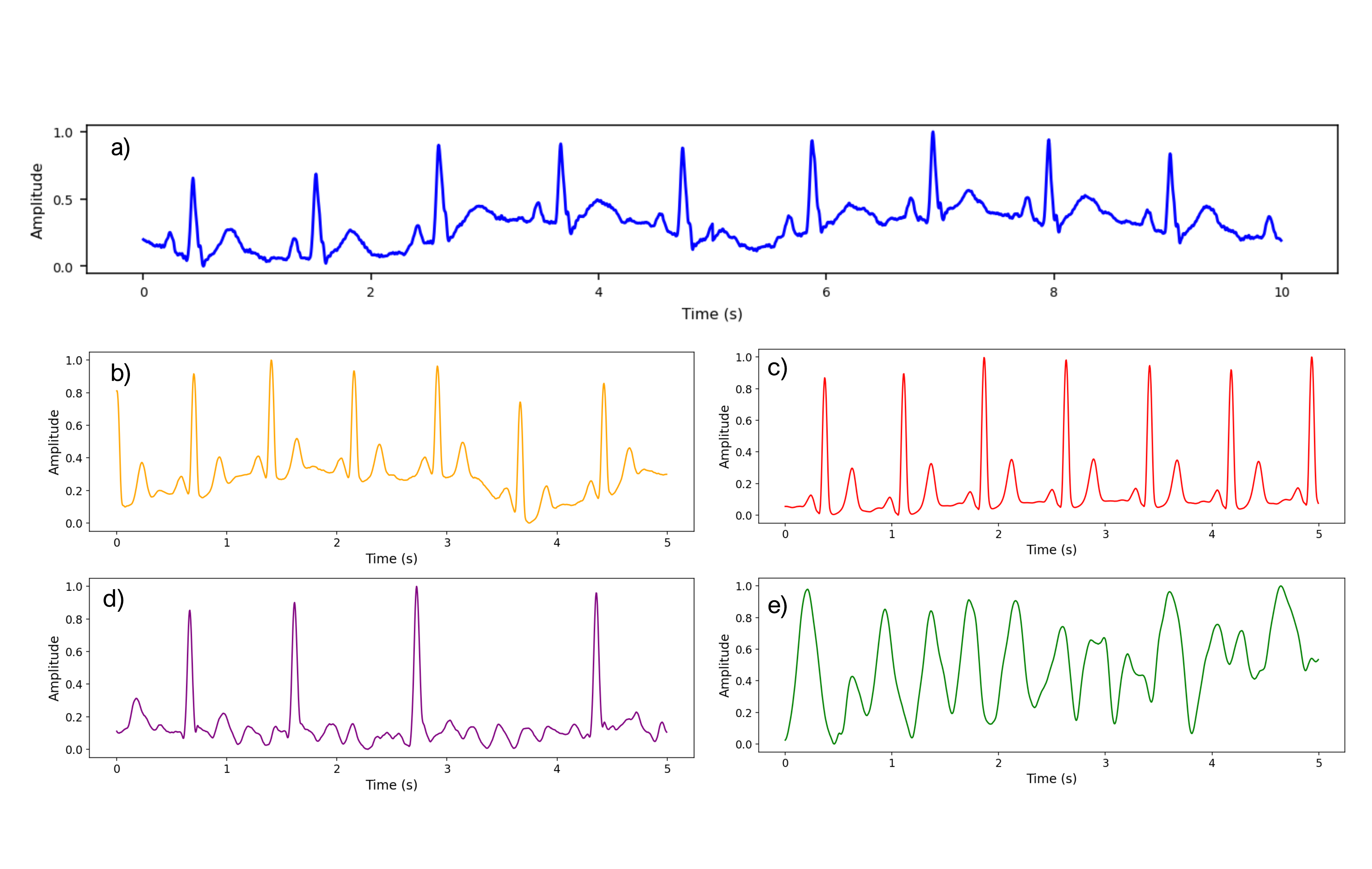} 
    \caption{ECG recording samples of 5 seconds of a) Normal Sinus Rhythm, b) Sinus Bradycardia, b) Sinus Tachycardia, c) Ventricular Flutter, and d) Atrial Fibrillation }
    \label{fig:ECGs}
\end{figure*}

Our final database contained 97 samples of Normal Sinus Rhythm (NSR), which serves as a standard reference of a healthy cardiac behavior. A sample from this pattern is displayed in Fig. \ref{fig:ECGs}a). Additionally, our database includes 100 samples of sinus bradycardia, 100 samples of sinus tachycardia, 24 samples of ventricular flutter, and 105 samples of atrial fibrillation.  Each sample was 5 seconds long and displayed a continuous arrhythmic pattern. A sample from each arrhythmia is displayed in figures \ref{fig:ECGs}b), \ref{fig:ECGs}c), \ref{fig:ECGs}d) and \ref{fig:ECGs}e).

\subsection{Signal preprocessing}

To standardize the data and ensure compatibility of all samples with the training phase, we performed thorough signal preprocessing. The MIT-BIH Arrhythmia Database ECG signals were processed using a downsampling algorithm to reduce the sampling frequency to 257 Hz,  matching the sampling frequency of the St. Petersburg INCART Database. 

Importantly, the signals were divided into 5-second segments. This process generated a dataset with 1,285 features. Each feature represents the electrical signal, measured in millivolts (mV). Subsequently, we applied a moving average algorithm to all samples with a window size of 12 samples to smooth the signals. This approach significantly reduced noise without compromising substantial information from the raw data. This final step led to a dataset with 1,273 features. Finally, the data was normalized to a range of 0 to 1 using the min-max scaling algorithm. 

\subsection{Model and architecture}

ML can be defined as a collection of algorithms capable of identifying intricate patterns within data and provide decision-making for specific tasks. As previously remarked, neural networks have gained prominence as a powerful and widely used ML algorithm approach. Neural networks consist of a series of interconnected nodes, or neurons, which perform a transformation. Each neuron computes an output by applying a nonlinear function, known as activation function, to a weighted sum of its inputs \cite{wu2018development}. During the training stage, the synaptic weights of each neuron are automatically adjusted through an optimization algorithm that minimizes a loss function. Finally, the performance is evaluated in the testing phase.Increasing the number of neurons in the network leads to higher computational cost, greater algorithmic complexity, and an elevated risk of overfitting. Overfitting is a common issue in supervised machine learning that hinders the creation of generalized models \cite{Ying_2019}. It occurs when the model becomes overly specialized in the training data. Consequently, the algorithm performs poorly when evaluated with unseen data, resulting in poor generalization. 

Our approach employs a single-hidden-layer architecture, substantially reducing the number of neurons in the network to lower computational costs and facilitate deployment in real-world applications. As a crucial part of our approach, we incorporated advanced feature engineering and signal analysis techniques to extract the maximum amount of relevant information from the raw ECG data. The combination of a simplified architecture and sophisticated data processing enables efficient and effective classification while minimizing resource requirements.

The input layer is configured to match the dimensionality of the feature vector, comprising seventeen characteristics. A batch normalization (BN) layer is then applied directly to its output to stabilize the learning process. The normalized data are subsequently passed to a hidden layer containing five neurons, each employing a "leaky\_relu" activation function. The output layer consists of five neurons, corresponding to the target classes, with each neuron using a softmax activation function to estimate the class probabilities. This architecture ensures a powerful and stable model while maintaining simplicity. 

The proposed model was trained and evaluated over 100 independent runs. In each run, the network was trained for 100 epochs with a learning rate of 0.03 and an $L_1$ regularization coefficient of $1 \times 10^{-4}$. The batch size was fixed at 36 samples, and the test set comprised 30\% of the dataset. The Adam optimizer was employed for stochastic optimization due to its computational efficiency and ease of implementation~\cite{kingma2014adam}.

\subsection{Feature Engineering}\label{subsec2}

To optimize the performance of the proposed neural network without increasing its architectural complexity, advanced feature engineering techniques were incorporated. The resulting feature vector comprised 17 key metrics derived from ECG signal data. These included the four classical statistical moments—mean, kurtosis, skewness, and variance~\cite{singh2023ecg}—as well as three clinically relevant features: cardiac rhythm, peak amplitude variability, and the average inter-peak interval. Collectively, these widely used features have been shown to be sufficient for classifying ECG signals with acceptable accuracy~\cite{sattar2023electrocardiogram}. In particular, the statistical moments have demonstrated relevance and reliability in capturing morphological characteristics of the data distribution and the underlying signal dynamics~\cite{141456,afkhami2016cardiac}. However, to achieve an accuracy exceeding 95\%, the proposed algorithm further integrates advanced feature engineering strategies.

Among these strategies, dimensionality reduction plays a crucial role in enhancing model efficiency while preserving essential information. Principal Component Analysis (PCA) is a statistical method that reduces the dimensionality of data. PCA is widely employed in signal studies to condense the data into key components that capture the most significant information and intricate characteristics. Interestingly, PCA has proven to be valuable in ECG analysis and signal processing. It is a particularly effective tool for extracting patterns and identifying physiological abnormalities. \cite{castells2007principal}. Additionally, researchers have employed PCA for the classification of arrhythmic diseases to reduce data dimensionality while preserving critical features \cite{378434}. For this work, we included three components derived from PCA as features for classification. 

Another component of our feature engineering strategy involved applying mathematical transformations to the ECG signals. As demonstrated by Sadhukhan \textit{et al.}~\cite{sadhukhan2018automated}, such transformations can extract discriminative features that enhance the interpretability of ECG data and substantially improve classification performance. Accordingly, we applied the Discrete Fourier Transform (DFT) to each ECG signal. The Fourier transform has been widely used in ECG analysis as a feature extraction tool, enabling models to achieve classification accuracies exceeding 97\%~\cite{9858160}. The DFT of a sequence \( x_0, x_1, \dots, x_{N-1} \) is defined as follows:

\begin{equation}
X_k = \sum_{n=0}^{N-1} x_n e^{-i 2 \pi \frac{k n}{N}},
\end{equation}

where \( X_k \) represents the \( k \)-th frequency component of the DFT, and \( x_n \) are the time-domain samples with \( n = 0, 1, \dots, N-1 \). Here, \( N \) is the number of samples in the sequence, and \( e^{-i 2 \pi \frac{k n}{N}} \) is a complex exponential function that represents the basis functions of the Fourier transform. In this context, \( i \) is the imaginary unit (\( i = \sqrt{-1} \)), and \( k = 0, 1, \dots, N-1 \) represents the frequency indices. The purpose of the DFT is to convert the original time-domain signal into its frequency components, which are represented by the complex values \( X_k \). These components contain the amplitude and phase information of the signal at different frequencies. After applying the algorithm, the Fourier transformation generated a set of complex numbers that represent the signal’s frequency components. This data was then processed to generate two derived features: skewness and kurtosis.

In addition to the DFT, we applied the Hilbert Transform, a widely used method in ECG signal studies and research. \cite{singh2023ecg, gupta2020efficient}. The purpose of the Hilbert Transform is to generate an enveloping function that captures the essential oscillatory behavior of ECG signals. The envelope generated from the Hilbert Transform is defined as a function of time applied to a a signal \( x(\tau) \) and is given by the following formula:

\begin{equation}
H\{x\}(t) =\frac{1}{\pi} \mathcal{\mathcal{P}}\int_{-\infty }^{\infty}\frac{x(\tau)}{t-\tau} d\tau
\end{equation}

where \( \mathcal{P} \) denotes the principal value of the integral, and \( t \) and \( \tau \) represent time variables \cite{benitez2001use}. After applying the transform, we processed the obtained signal to extract the mean, minimum amplitude, and maximum amplitude from the resulting function. 

Another key feature in our methodology is entropy, a metric quantifying a signal’s unpredictability, disorder, and complexity. In ECG signal analysis, entropy has proven to be a valuable feature, particularly for diagnosing atrial fibrillation, one of the most prevalent arrhythmic disorders~\cite{kumar2018automated,asgharzadeh2020spectral,a2021three}. Our approach incorporates Shannon entropy, which measures the complexity of data and is defined as follows:

\begin{equation}
H(X) = - \sum_{i=1}^{n} p(x_i) \log_2 p(x_i),
\end{equation}

where \( H(X) \) represents the Shannon entropy of the  \( X \), \( p(x_i) \) is the probability mass function of the \( i \)-th outcome \( x_i \), \( \log_2 \) denotes the logarithm to the base 2, \( n \) is the number of possible outcomes for the random variable \( X \), and \( x_i \) represents the \( i \)-th outcome of the random variable. Shannon entropy was computed for each 5-second sample. 

Additional features include heart rate, defined as the number of heartbeats per unit time, typically expressed in beats per minute (bpm). The heartbeat frequency corresponds to the cardiac cycle rate and is determined by the time intervals between consecutive peaks in the ECG signal. Another important feature is peak amplitude variability, quantified as the standard deviation of amplitude differences between successive peaks. Lastly, the average distance between peaks represents the mean elapsed time between consecutive heartbeats, providing insights into the cardiac rhythm’s behavior. These features are widely employed in cardiology research for diagnosing arrhythmic disorders and are straightforward to compute~\cite{singh2023ecg}.

A more detailed description of similar features can be found in Frausto et al.~\cite{frausto2025classification}, where the authors proposed a comparable feature vector to classify SERS spectra of organophosphate pesticides. Collectively, these 17 metrics form a comprehensive feature vector that captures relevant signal characteristics, enabling the neural network to maximize performance while maintaining low algorithmic complexity.

We advocate the use of these features to extract significant, easily computed descriptors that effectively characterize the ECG signal’s behavior and key morphological traits, thereby eliminating the need for deeper neural network architectures.

\section{Results}\label{sec3}

\begin{figure}
    \centering
    \includegraphics[width=9cm]{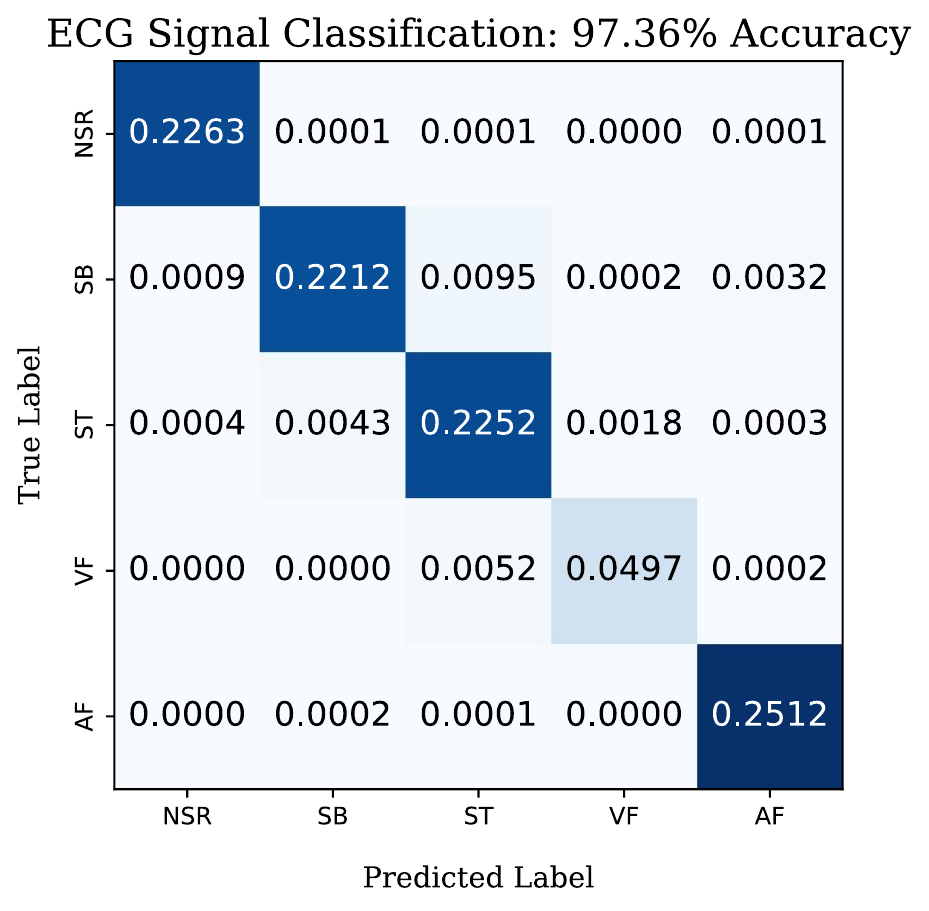}
    \caption{ Confusion matrix of the neural network performance}
    \label{fig:PQRST_architecture}
\end{figure}

The results were recorded using a confusion matrix after each iteration. The outcomes from 100 iterations were averaged to generate the final confusion matrix which is presented in Fig. \ref{fig:PQRST_architecture}. Our compact neural network algorithm has achieved an exceptionally strong performance, reaching an accuracy of 97.36\%. These promising results indicate that the features extracted from the ECG signals provide a reliable, comprehensive and relevant data source for the detection and classification of arrhythmic disorders. Further analysis of the results, including sensitivity, specificity, F1 score, and accuracy for each class, is documented in Table~\ref{tab:results}.

\begin{table}[t!]
    \centering
    \caption{Performance metrics of the classification algorithm for each class. } 
    \begin{tabular*}{\linewidth}{@{\extracolsep{\fill}}c c c c}
    \hline
    \textbf{Class} & \textbf{Precision} & \textbf{Recall} & \textbf{F1 Score} \\ 
    \midrule
    NSR   & 0.99      & 1          & 1         \\ 
    SB    & 0.98      & 0.94         & 0.96    \\ 
    ST    & 0.94      & 0.97         & 0.95     \\ 
    VF    & 0.96      & 0.9         & 0.93     \\ 
    AF    & 0.99      & 1            & 0.99    \\ 
    \bottomrule
    \end{tabular*}
    \label{tab:results}
\end{table}

\begin{sidewaystable*}[t!]
\centering
\caption{Performance metrics of other state-of-the-art models.} 
\begin{tabular*}{\linewidth}{@{\extracolsep{\fill}}l l l l l }

\toprule
\textbf{Type of network} & \textbf{Accuracy (\%)} & \textbf{Year} & \textbf{Author} & \textbf{Arrythmic disorders} \\
\midrule

\begin{tabular}[c]{@{}l@{}}Compact Artificial\\ Neural Network (ANN)\end{tabular} & 97.36    & Current work & Ours     & \begin{tabular}[c]{@{}l@{}}Sinus Bradycardia\\ Sinus Tachycardia\\ Ventricular Flutter \\ Atrial Fibrillation\\ Normal Sinus Rhythm\end{tabular}   

\\ \hline
Self-ONN                                                                          & 82, 94   & 2024         & {[}36{]} & \begin{tabular}[c]{@{}l@{}}Supraventricular Ectopic Beats (SVEPs)\\ Ventricular Ectopic Beats (VEPs)\end{tabular}                                                                                              \\ \hline
12-layer CNN                                                                      & 97.8     & 2024         & {[}37{]} & \begin{tabular}[c]{@{}l@{}}Left Bundle Branch Block (LBBB)\\ Right Bundle Brack Block (RBBB)\\ Atrial Premature Contraction (APC)\\ Premature Ventricular Contraction (PVC)\\ Normal Sinus Rhythm\end{tabular} \\ \hline
1-D CNN                                                                           & 98.63    & 2023         & {[}38{]} & \begin{tabular}[c]{@{}l@{}}SVEPs\\ VEPs\\ Fusion Beat\\ Normal Sinus Rhythm\end{tabular}                                                                                                                       \\ \hline
2-D CNN                                                                           & 96.04    & 2018         & {[}16{]} & \begin{tabular}[c]{@{}l@{}}SVEPs\\ VEPs\\ Fusion Beat\\ Normla Sinus Rhythm\end{tabular}                                                                                                                       \\ \hline
CNN                                                                               & 92.5     & 2017         & {[}39{]} & \begin{tabular}[c]{@{}l@{}}Sinus Tachycardia\\ Atrial Fibrillation\\ Atrial Flutter\\ Normal Sinus Rhythm\end{tabular}                                                                                         \\ \hline
Decision Tree                                                                     & 96.3     & 2016         & {[}21{]} & \begin{tabular}[c]{@{}l@{}}Atrial Fibrillation\\ Atrial Flutter\\ Ventricular Flutter\\ Normal Sinus Rhythm\end{tabular}                                                                                       \\ \bottomrule
\end{tabular*}
\label{tab:compa}
\end{sidewaystable*}

The achieved performance of our model is comparable to that of other state-of-the-art models. To contextualize our results, we compare our approach with several relevant studies from the literature. Ref. \cite{sahoo2017multiresolution} introduced a neural network incorporating wavelet transform-based engineered features, achieving an accuracy of 96.67\%. In Ref. \cite{10563956} developed a relatively compact and low computationally complex 1-D Self-Operational Neural Network (Self-ONN) for ECG classification.  Nonetheless, this model was limited to classifying ectopic and ventricular ectopic beats, achieving 82\% and 94\% accuracy, respectively. On the other hand, Ref. \cite{challagundla2024advanced} presented a complex 12-layer CNN for classifying  arrhythmic disorders similar to those used in this study. The model achieved a slightly higher accuracy of 97.8\%. Nevertheless, this performance came at the cost of a significantly more complex architecture, which increases computational complexity. In another approach, Ref. \cite{khan2023ecg} implemented a 64-filter, 1-D CNN with a superior average accuracy of 98.63\%. However, our algorithm offers a comparable performance with lower computational complexity, making it more suitable for real-world applications that often face resource-limited environments. In Ref. \cite{acharya2017automated} presented a CNN model that classified similar disorders to those used in this study with an accuracy of 92.5\%. In a separate study, \cite{acharya2016automated} proposed a decision tree-based approach that achieved an accuracy of 96.3\%. As demonstrated, our algorithm performs similarly to more advanced models, including CNN-based approaches. However, it achieves comparable accuracy while maintaining minimal computational complexity. A  complete comparison of our model with other state-of-the-art algorithms is presented in Table~\ref{tab:compa}.

\section{Conclusions}\label{sec4}

Our research makes a significant contribution to the study of arrhythmic disorders through ECG signal analysis and machine learning algorithms. We have demonstrated that compact neural networks offer a robust, ML-based approach for ECG classification tasks. Notably, the simplicity of the proposed neural network architecture enables straightforward software implementation, facilitating integration into diverse healthcare settings, including those with limited technological resources. Beyond the neural network model itself, the proposed feature engineering techniques allow us to extract relevant and intricate information from ECG signals. The performance of our model is comparable to that of state-of-the-art algorithms. We encourage further exploration of feature engineering as an effective strategy to develop simpler, more accessible, and robust machine learning solutions for arrhythmic disorder detection and classification.

\section*{Acknowledgments}

\noindent M.A.Q.-J. thankfully acknowledges financial support by SECIHTI/CONAHCyT under the Project CF-2023-I-1496 and by DGAPA-UNAM under the Project UNAM-PAPIIT IA103325. A.B.U. thankfully acknowledges financial support by DGAPA-UNAM under the project UNAM-PAPIIT IN103521. C.M.F.A and A.K.S.R.R. thankfully acknowledges financial support by SECIHTI/CONAHCyT for posdoctoral grant.

\section*{Competing interests}
\noindent The authors declare no competing interests.

\section*{Data and Code Availability}
The source code for this study is available at: \url{https://github.com/CFATA-AI/ECG-CLASSIFICATION}. The research utilized two PhysioNet databases: MIT-BIH Arrhythmia Database (\url{https://www.physionet.org/content/mitdb/1.0.0}) and INCART DB (\url{https://physionet.org/content/incartdb/1.0.0}).





\begin{thebibliography}{47}
\ifx \bisbn   \undefined \def \bisbn  #1{ISBN #1}\fi
\ifx \binits  \undefined \def \binits#1{#1}\fi
\ifx \bauthor  \undefined \def \bauthor#1{#1}\fi
\ifx \batitle  \undefined \def \batitle#1{#1}\fi
\ifx \bjtitle  \undefined \def \bjtitle#1{#1}\fi
\ifx \bvolume  \undefined \def \bvolume#1{\textbf{#1}}\fi
\ifx \byear  \undefined \def \byear#1{#1}\fi
\ifx \bissue  \undefined \def \bissue#1{#1}\fi
\ifx \bfpage  \undefined \def \bfpage#1{#1}\fi
\ifx \blpage  \undefined \def \blpage #1{#1}\fi
\ifx \burl  \undefined \def \burl#1{\textsf{#1}}\fi
\ifx \doiurl  \undefined \def \doiurl#1{\url{https://doi.org/#1}}\fi
\ifx \betal  \undefined \def \betal{\textit{et al.}}\fi
\ifx \binstitute  \undefined \def \binstitute#1{#1}\fi
\ifx \binstitutionaled  \undefined \def \binstitutionaled#1{#1}\fi
\ifx \bctitle  \undefined \def \bctitle#1{#1}\fi
\ifx \beditor  \undefined \def \beditor#1{#1}\fi
\ifx \bpublisher  \undefined \def \bpublisher#1{#1}\fi
\ifx \bbtitle  \undefined \def \bbtitle#1{#1}\fi
\ifx \bedition  \undefined \def \bedition#1{#1}\fi
\ifx \bseriesno  \undefined \def \bseriesno#1{#1}\fi
\ifx \blocation  \undefined \def \blocation#1{#1}\fi
\ifx \bsertitle  \undefined \def \bsertitle#1{#1}\fi
\ifx \bsnm \undefined \def \bsnm#1{#1}\fi
\ifx \bsuffix \undefined \def \bsuffix#1{#1}\fi
\ifx \bparticle \undefined \def \bparticle#1{#1}\fi
\ifx \barticle \undefined \def \barticle#1{#1}\fi
\bibcommenthead
\ifx \bconfdate \undefined \def \bconfdate #1{#1}\fi
\ifx \botherref \undefined \def \botherref #1{#1}\fi
\ifx \url \undefined \def \url#1{\textsf{#1}}\fi
\ifx \bchapter \undefined \def \bchapter#1{#1}\fi
\ifx \bbook \undefined \def \bbook#1{#1}\fi
\ifx \bcomment \undefined \def \bcomment#1{#1}\fi
\ifx \oauthor \undefined \def \oauthor#1{#1}\fi
\ifx \citeauthoryear \undefined \def \citeauthoryear#1{#1}\fi
\ifx \endbibitem  \undefined \def \endbibitem {}\fi
\ifx \bconflocation  \undefined \def \bconflocation#1{#1}\fi
\ifx \arxivurl  \undefined \def \arxivurl#1{\textsf{#1}}\fi
\csname PreBibitemsHook\endcsname

\bibitem[\protect\citeauthoryear{Singh and Krishnan}{2023}]{singh2023ecg}
\begin{barticle}
\bauthor{\bsnm{Singh}, \binits{A.K.}},
\bauthor{\bsnm{Krishnan}, \binits{S.}}:
\batitle{Ecg signal feature extraction trends in methods and applications}.
\bjtitle{BioMedical Engineering OnLine}
\bvolume{22}(\bissue{1}),
\bfpage{22}
(\byear{2023})
\end{barticle}
\endbibitem

\bibitem[\protect\citeauthoryear{Mendis et~al.}{2015}]{mendis2015organizational}
\begin{barticle}
\bauthor{\bsnm{Mendis}, \binits{S.}},
\bauthor{\bsnm{Davis}, \binits{S.}},
\bauthor{\bsnm{Norrving}, \binits{B.}}:
\batitle{Organizational update: the world health organization global status report on noncommunicable diseases 2014; one more landmark step in the combat against stroke and vascular disease}.
\bjtitle{Stroke}
\bvolume{46}(\bissue{5}),
\bfpage{121}--\blpage{122}
(\byear{2015})
\end{barticle}
\endbibitem

\bibitem[\protect\citeauthoryear{AlGhatrif and Lindsay}{2012}]{alghatrif2012brief}
\begin{barticle}
\bauthor{\bsnm{AlGhatrif}, \binits{M.}},
\bauthor{\bsnm{Lindsay}, \binits{J.}}:
\batitle{A brief review: history to understand fundamentals of electrocardiography}.
\bjtitle{Journal of community hospital internal medicine perspectives}
\bvolume{2}(\bissue{1}),
\bfpage{14383}
(\byear{2012})
\end{barticle}
\endbibitem

\bibitem[\protect\citeauthoryear{Sattar and Chhabra}{2023}]{sattar2023electrocardiogram}
\begin{bchapter}
\bauthor{\bsnm{Sattar}, \binits{Y.}},
\bauthor{\bsnm{Chhabra}, \binits{L.}}:
\bctitle{Electrocardiogram}.
In: \bbtitle{StatPearls [Internet]}.
\bpublisher{StatPearls Publishing}, \blocation{???}
(\byear{2023})
\end{bchapter}
\endbibitem

\bibitem[\protect\citeauthoryear{Quiroz-Ju{\'a}rez et~al.}{2019}]{quiroz2019periodically}
\begin{barticle}
\bauthor{\bsnm{Quiroz-Ju{\'a}rez}, \binits{M.A.}},
\bauthor{\bsnm{Jim{\'e}nez-Ram{\'\i}rez}, \binits{O.}},
\bauthor{\bsnm{Arag{\'o}n}, \binits{J.}},
\bauthor{\bsnm{Del~R{\'\i}o-Correa}, \binits{J.L.}},
\bauthor{\bsnm{V{\'a}zquez-Medina}, \binits{R.}}:
\batitle{Periodically kicked network of rlc oscillators to produce ecg signals}.
\bjtitle{Computers in biology and medicine}
\bvolume{104},
\bfpage{87}--\blpage{96}
(\byear{2019})
\end{barticle}
\endbibitem

\bibitem[\protect\citeauthoryear{Gois and Savi}{2009}]{gois2009analysis}
\begin{barticle}
\bauthor{\bsnm{Gois}, \binits{S.R.}},
\bauthor{\bsnm{Savi}, \binits{M.A.}}:
\batitle{An analysis of heart rhythm dynamics using a three-coupled oscillator model}.
\bjtitle{Chaos, Solitons \& Fractals}
\bvolume{41}(\bissue{5}),
\bfpage{2553}--\blpage{2565}
(\byear{2009})
\end{barticle}
\endbibitem

\bibitem[\protect\citeauthoryear{Quiroz-Ju{\'a}rez et~al.}{2019}]{quiroz2019generation}
\begin{barticle}
\bauthor{\bsnm{Quiroz-Ju{\'a}rez}, \binits{M.}},
\bauthor{\bsnm{Jim{\'e}nez-Ram{\'\i}rez}, \binits{O.}},
\bauthor{\bsnm{V{\'a}zquez-Medina}, \binits{R.}},
\bauthor{\bsnm{Bre{\~n}a-Medina}, \binits{V.}},
\bauthor{\bsnm{Arag{\'o}n}, \binits{J.}},
\bauthor{\bsnm{Barrio}, \binits{R.}}:
\batitle{Generation of ecg signals from a reaction-diffusion model spatially discretized}.
\bjtitle{Scientific reports}
\bvolume{9}(\bissue{1}),
\bfpage{19000}
(\byear{2019})
\end{barticle}
\endbibitem

\bibitem[\protect\citeauthoryear{McSharry et~al.}{2003}]{mcsharry2003dynamical}
\begin{barticle}
\bauthor{\bsnm{McSharry}, \binits{P.E.}},
\bauthor{\bsnm{Clifford}, \binits{G.D.}},
\bauthor{\bsnm{Tarassenko}, \binits{L.}},
\bauthor{\bsnm{Smith}, \binits{L.A.}}:
\batitle{A dynamical model for generating synthetic electrocardiogram signals}.
\bjtitle{IEEE transactions on biomedical engineering}
\bvolume{50}(\bissue{3}),
\bfpage{289}--\blpage{294}
(\byear{2003})
\end{barticle}
\endbibitem

\bibitem[\protect\citeauthoryear{Quiroz-Juarez et~al.}{2018}]{quiroz2018cardiac}
\begin{barticle}
\bauthor{\bsnm{Quiroz-Juarez}, \binits{M.A.}},
\bauthor{\bsnm{Jimenez-Ramirez}, \binits{O.}},
\bauthor{\bsnm{Vazquez-Medina}, \binits{R.}},
\bauthor{\bsnm{Ryzhii}, \binits{E.}},
\bauthor{\bsnm{Ryzhii}, \binits{M.}},
\bauthor{\bsnm{Aragon}, \binits{J.L.}}:
\batitle{Cardiac conduction model for generating 12 lead ecg signals with realistic heart rate dynamics}.
\bjtitle{IEEE transactions on nanobioscience}
\bvolume{17}(\bissue{4}),
\bfpage{525}--\blpage{532}
(\byear{2018})
\end{barticle}
\endbibitem

\bibitem[\protect\citeauthoryear{Khan et~al.}{2023}]{khan2023ecg}
\begin{barticle}
\bauthor{\bsnm{Khan}, \binits{F.}},
\bauthor{\bsnm{Yu}, \binits{X.}},
\bauthor{\bsnm{Yuan}, \binits{Z.}},
\bauthor{\bsnm{Rehman}, \binits{A.U.}}:
\batitle{Ecg classification using 1-d convolutional deep residual neural network}.
\bjtitle{Plos one}
\bvolume{18}(\bissue{4}),
\bfpage{0284791}
(\byear{2023})
\end{barticle}
\endbibitem

\bibitem[\protect\citeauthoryear{Y{\i}ld{\i}r{\i}m et~al.}{2018}]{yildirim2018arrhythmia}
\begin{barticle}
\bauthor{\bsnm{Y{\i}ld{\i}r{\i}m}, \binits{{\"O}.}},
\bauthor{\bsnm{P{\l}awiak}, \binits{P.}},
\bauthor{\bsnm{Tan}, \binits{R.-S.}},
\bauthor{\bsnm{Acharya}, \binits{U.R.}}:
\batitle{Arrhythmia detection using deep convolutional neural network with long duration ecg signals}.
\bjtitle{Computers in biology and medicine}
\bvolume{102},
\bfpage{411}--\blpage{420}
(\byear{2018})
\end{barticle}
\endbibitem

\bibitem[\protect\citeauthoryear{Mjolsness and DeCoste}{2001}]{ML_Intr_1}
\begin{barticle}
\bauthor{\bsnm{Mjolsness}, \binits{E.}},
\bauthor{\bsnm{DeCoste}, \binits{D.}}:
\batitle{Machine learning for science: state of the art and future prospects}.
\bjtitle{science}
\bvolume{293}(\bissue{5537}),
\bfpage{2051}--\blpage{2055}
(\byear{2001})
\end{barticle}
\endbibitem

\bibitem[\protect\citeauthoryear{Villegas et~al.}{2022}]{villegas2022identification}
\begin{bchapter}
\bauthor{\bsnm{Villegas}, \binits{A.}},
\bauthor{\bsnm{Quiroz-Ju{\'a}rez}, \binits{M.A.}},
\bauthor{\bsnm{U’Ren}, \binits{A.B.}},
\bauthor{\bsnm{Torres}, \binits{J.P.}},
\bauthor{\bsnm{Le{\'o}n-Montiel}, \binits{R.d.J.}}:
\bctitle{Identification of model particle mixtures using machine-learning-assisted laser diffraction}.
In: \bbtitle{Photonics},
vol. \bseriesno{9},
p. \bfpage{74}
(\byear{2022}).
\bcomment{MDPI}
\end{bchapter}
\endbibitem

\bibitem[\protect\citeauthoryear{Lollie et~al.}{2022}]{lollie2022high}
\begin{barticle}
\bauthor{\bsnm{Lollie}, \binits{M.L.}},
\bauthor{\bsnm{Mostafavi}, \binits{F.}},
\bauthor{\bsnm{Bhusal}, \binits{N.}},
\bauthor{\bsnm{Hong}, \binits{M.}},
\bauthor{\bsnm{You}, \binits{C.}},
\bauthor{\bsnm{J~Le{\'o}n-Montiel}, \binits{R.}},
\bauthor{\bsnm{Maga{\~n}a-Loaiza}, \binits{O.S.}},
\bauthor{\bsnm{Quiroz-Ju{\'a}rez}, \binits{M.A.}}:
\batitle{High-dimensional encryption in optical fibers using spatial modes of light and machine learning}.
\bjtitle{Machine Learning: Science and Technology}
\bvolume{3}(\bissue{3}),
\bfpage{035006}
(\byear{2022})
\end{barticle}
\endbibitem

\bibitem[\protect\citeauthoryear{Salazar et~al.}{2024}]{salazar2024improving}
\begin{barticle}
\bauthor{\bsnm{Salazar}, \binits{M.F.}},
\bauthor{\bsnm{Frausto-Avila}, \binits{C.M.}},
\bauthor{\bsnm{Jes{\'u}s~Bautista}, \binits{J.A.}},
\bauthor{\bsnm{Polumati}, \binits{G.}},
\bauthor{\bsnm{Mart{\'\i}nez}, \binits{B.A.M.}},
\bauthor{\bsnm{Reddy}, \binits{K.C.S.}},
\bauthor{\bsnm{Hern{\'a}ndez-V{\'a}zquez}, \binits{M.{\'A}.}},
\bauthor{\bsnm{Strupiechonski}, \binits{E.}},
\bauthor{\bsnm{Sahatiya}, \binits{P.}},
\bauthor{\bsnm{Quiroz-Ju{\'a}rez}, \binits{M.A.}}, \betal:
\batitle{Improving the coverage area and flake size of res2 through machine learning in apcvd}.
\bjtitle{Nanotechnology}
\bvolume{35}(\bissue{50}),
\bfpage{505705}
(\byear{2024})
\end{barticle}
\endbibitem

\bibitem[\protect\citeauthoryear{Zhai and Tin}{2018}]{8356203}
\begin{barticle}
\bauthor{\bsnm{Zhai}, \binits{X.}},
\bauthor{\bsnm{Tin}, \binits{C.}}:
\batitle{Automated ecg classification using dual heartbeat coupling based on convolutional neural network}.
\bjtitle{IEEE Access}
\bvolume{6},
\bfpage{27465}--\blpage{27472}
(\byear{2018})
\doiurl{10.1109/ACCESS.2018.2833841}
\end{barticle}
\endbibitem

\bibitem[\protect\citeauthoryear{Jun et~al.}{2018}]{jun2018ecg}
\begin{botherref}
\oauthor{\bsnm{Jun}, \binits{T.J.}},
\oauthor{\bsnm{Nguyen}, \binits{H.M.}},
\oauthor{\bsnm{Kang}, \binits{D.}},
\oauthor{\bsnm{Kim}, \binits{D.}},
\oauthor{\bsnm{Kim}, \binits{D.}},
\oauthor{\bsnm{Kim}, \binits{Y.-H.}}:
Ecg arrhythmia classification using a 2-d convolutional neural network.
arXiv preprint arXiv:1804.06812
(2018)
\end{botherref}
\endbibitem

\bibitem[\protect\citeauthoryear{Wang et~al.}{2021}]{wang2021automatic}
\begin{barticle}
\bauthor{\bsnm{Wang}, \binits{T.}},
\bauthor{\bsnm{Lu}, \binits{C.}},
\bauthor{\bsnm{Sun}, \binits{Y.}},
\bauthor{\bsnm{Yang}, \binits{M.}},
\bauthor{\bsnm{Liu}, \binits{C.}},
\bauthor{\bsnm{Ou}, \binits{C.}}:
\batitle{Automatic ecg classification using continuous wavelet transform and convolutional neural network}.
\bjtitle{Entropy}
\bvolume{23}(\bissue{1}),
\bfpage{119}
(\byear{2021})
\end{barticle}
\endbibitem

\bibitem[\protect\citeauthoryear{Sadhukhan et~al.}{2018}]{sadhukhan2018automated}
\begin{barticle}
\bauthor{\bsnm{Sadhukhan}, \binits{D.}},
\bauthor{\bsnm{Pal}, \binits{S.}},
\bauthor{\bsnm{Mitra}, \binits{M.}}:
\batitle{Automated identification of myocardial infarction using harmonic phase distribution pattern of ecg data}.
\bjtitle{IEEE Transactions on Instrumentation and Measurement}
\bvolume{67}(\bissue{10}),
\bfpage{2303}--\blpage{2313}
(\byear{2018})
\end{barticle}
\endbibitem

\bibitem[\protect\citeauthoryear{Zeng et~al.}{2024}]{zeng2024detection}
\begin{botherref}
\oauthor{\bsnm{Zeng}, \binits{W.}},
\oauthor{\bsnm{Shan}, \binits{L.}},
\oauthor{\bsnm{Yuan}, \binits{C.}},
\oauthor{\bsnm{Du}, \binits{S.}}:
Detection of myocardial infarction using shannon energy envelope, fa-mvemd and deterministic learning.
Complex \& Intelligent Systems,
1--19
(2024)
\end{botherref}
\endbibitem

\bibitem[\protect\citeauthoryear{Rabee and Barhumi}{2012}]{SVM_ECG}
\begin{bchapter}
\bauthor{\bsnm{Rabee}, \binits{A.}},
\bauthor{\bsnm{Barhumi}, \binits{I.}}:
\bctitle{Ecg signal classification using support vector machine based on wavelet multiresolution analysis}.
In: \bbtitle{2012 11th International Conference on Information Science, Signal Processing and Their Applications (ISSPA)},
pp. \bfpage{1319}--\blpage{1323}
(\byear{2012}).
\bcomment{IEEE}
\end{bchapter}
\endbibitem

\bibitem[\protect\citeauthoryear{Kumari et~al.}{2022}]{DT_ECG}
\begin{barticle}
\bauthor{\bsnm{Kumari}, \binits{L.}},
\bauthor{\bsnm{Sai}, \binits{Y.P.}}, \betal:
\batitle{Classification of ecg beats using optimized decision tree and adaptive boosted optimized decision tree}.
\bjtitle{Signal, Image and Video Processing}
\bvolume{16}(\bissue{3}),
\bfpage{695}--\blpage{703}
(\byear{2022})
\end{barticle}
\endbibitem

\bibitem[\protect\citeauthoryear{Mohebbanaaz et~al.}{2021}]{KNN_ECG}
\begin{bchapter}
\bauthor{\bsnm{Mohebbanaaz}},
\bauthor{\bsnm{Rajani~Kumari}, \binits{L.}},
\bauthor{\bsnm{Padma~Sai}, \binits{Y.}}:
\bctitle{Classification of arrhythmia beats using optimized k-nearest neighbor classifier}.
In: \bbtitle{Intelligent Systems: Proceedings of ICMIB 2020},
pp. \bfpage{349}--\blpage{359}
(\byear{2021}).
\bcomment{Springer}
\end{bchapter}
\endbibitem

\bibitem[\protect\citeauthoryear{Ren et~al.}{2024}]{ren2024exploring}
\begin{barticle}
\bauthor{\bsnm{Ren}, \binits{Z.}},
\bauthor{\bsnm{Lan}, \binits{Q.}},
\bauthor{\bsnm{Zhang}, \binits{Y.}},
\bauthor{\bsnm{Wang}, \binits{S.}}:
\batitle{Exploring simple triplet representation learning}.
\bjtitle{Computational and Structural Biotechnology Journal}
\bvolume{23},
\bfpage{1510}--\blpage{1521}
(\byear{2024})
\end{barticle}
\endbibitem

\bibitem[\protect\citeauthoryear{Ren et~al.}{2022}]{ren2022hybrid}
\begin{barticle}
\bauthor{\bsnm{Ren}, \binits{Z.}},
\bauthor{\bsnm{Zhang}, \binits{Y.}},
\bauthor{\bsnm{Wang}, \binits{S.}}:
\batitle{A hybrid framework for lung cancer classification}.
\bjtitle{Electronics}
\bvolume{11}(\bissue{10}),
\bfpage{1614}
(\byear{2022})
\end{barticle}
\endbibitem

\bibitem[\protect\citeauthoryear{Zhou et~al.}{2024}]{foundation_cancer_segmentation}
\begin{barticle}
\bauthor{\bsnm{Zhou}, \binits{Y.}},
\bauthor{\bsnm{Wang}, \binits{L.}},
\bauthor{\bsnm{Kim}, \binits{H.}}:
\batitle{Large foundation model for cancer segmentation}.
\bjtitle{Nature Machine Intelligence}
\bvolume{6}(\bissue{1}),
\bfpage{45}--\blpage{58}
(\byear{2024})
\doiurl{10.1038/s41524-024-00234-9}
\end{barticle}
\endbibitem

\bibitem[\protect\citeauthoryear{Moody and Mark}{2001}]{moody2001impact}
\begin{barticle}
\bauthor{\bsnm{Moody}, \binits{G.B.}},
\bauthor{\bsnm{Mark}, \binits{R.G.}}:
\batitle{The impact of the mit-bih arrhythmia database}.
\bjtitle{IEEE engineering in medicine and biology magazine}
\bvolume{20}(\bissue{3}),
\bfpage{45}--\blpage{50}
(\byear{2001})
\end{barticle}
\endbibitem

\bibitem[\protect\citeauthoryear{Goldberger et~al.}{2000}]{goldberger2000physiobank}
\begin{barticle}
\bauthor{\bsnm{Goldberger}, \binits{A.L.}},
\bauthor{\bsnm{Amaral}, \binits{L.A.}},
\bauthor{\bsnm{Glass}, \binits{L.}},
\bauthor{\bsnm{Hausdorff}, \binits{J.M.}},
\bauthor{\bsnm{Ivanov}, \binits{P.C.}},
\bauthor{\bsnm{Mark}, \binits{R.G.}},
\bauthor{\bsnm{Mietus}, \binits{J.E.}},
\bauthor{\bsnm{Moody}, \binits{G.B.}},
\bauthor{\bsnm{Peng}, \binits{C.-K.}},
\bauthor{\bsnm{Stanley}, \binits{H.E.}}:
\batitle{Physiobank, physiotoolkit, and physionet: components of a new research resource for complex physiologic signals}.
\bjtitle{circulation}
\bvolume{101}(\bissue{23}),
\bfpage{215}--\blpage{220}
(\byear{2000})
\end{barticle}
\endbibitem

\bibitem[\protect\citeauthoryear{Wu and Feng}{2018}]{wu2018development}
\begin{barticle}
\bauthor{\bsnm{Wu}, \binits{Y.-c.}},
\bauthor{\bsnm{Feng}, \binits{J.-w.}}:
\batitle{Development and application of artificial neural network}.
\bjtitle{Wireless Personal Communications}
\bvolume{102},
\bfpage{1645}--\blpage{1656}
(\byear{2018})
\end{barticle}
\endbibitem

\bibitem[\protect\citeauthoryear{Ying}{2019}]{Ying_2019}
\begin{barticle}
\bauthor{\bsnm{Ying}, \binits{X.}}:
\batitle{An overview of overfitting and its solutions}.
\bjtitle{Journal of Physics: Conference Series}
\bvolume{1168}(\bissue{2}),
\bfpage{022022}
(\byear{2019})
\doiurl{10.1088/1742-6596/1168/2/022022}
\end{barticle}
\endbibitem

\bibitem[\protect\citeauthoryear{Kingma}{2014}]{kingma2014adam}
\begin{botherref}
\oauthor{\bsnm{Kingma}, \binits{D.P.}}:
Adam: A method for stochastic optimization.
arXiv preprint arXiv:1412.6980
(2014)
\end{botherref}
\endbibitem

\bibitem[\protect\citeauthoryear{Soliman and Hsue}{1992}]{141456}
\begin{barticle}
\bauthor{\bsnm{Soliman}, \binits{S.S.}},
\bauthor{\bsnm{Hsue}, \binits{S.-Z.}}:
\batitle{Signal classification using statistical moments}.
\bjtitle{IEEE Transactions on Communications}
\bvolume{40}(\bissue{5}),
\bfpage{908}--\blpage{916}
(\byear{1992})
\doiurl{10.1109/26.141456}
\end{barticle}
\endbibitem

\bibitem[\protect\citeauthoryear{Afkhami et~al.}{2016}]{afkhami2016cardiac}
\begin{barticle}
\bauthor{\bsnm{Afkhami}, \binits{R.G.}},
\bauthor{\bsnm{Azarnia}, \binits{G.}},
\bauthor{\bsnm{Tinati}, \binits{M.A.}}:
\batitle{Cardiac arrhythmia classification using statistical and mixture modeling features of ecg signals}.
\bjtitle{Pattern Recognition Letters}
\bvolume{70},
\bfpage{45}--\blpage{51}
(\byear{2016})
\end{barticle}
\endbibitem

\bibitem[\protect\citeauthoryear{Castells et~al.}{2007}]{castells2007principal}
\begin{barticle}
\bauthor{\bsnm{Castells}, \binits{F.}},
\bauthor{\bsnm{Laguna}, \binits{P.}},
\bauthor{\bsnm{S{\"o}rnmo}, \binits{L.}},
\bauthor{\bsnm{Bollmann}, \binits{A.}},
\bauthor{\bsnm{Roig}, \binits{J.M.}}:
\batitle{Principal component analysis in ecg signal processing}.
\bjtitle{EURASIP Journal on Advances in Signal Processing}
\bvolume{2007},
\bfpage{1}--\blpage{21}
(\byear{2007})
\end{barticle}
\endbibitem

\bibitem[\protect\citeauthoryear{Nadal and de~C.~Bossan}{1993}]{378434}
\begin{bchapter}
\bauthor{\bsnm{Nadal}, \binits{J.}},
\bauthor{\bsnm{C.~Bossan}, \binits{M.}}:
\bctitle{Classification of cardiac arrhythmias based on principal component analysis and feedforward neural networks}.
In: \bbtitle{Proceedings of Computers in Cardiology Conference},
pp. \bfpage{341}--\blpage{344}
(\byear{1993}).
\doiurl{10.1109/CIC.1993.378434}
\end{bchapter}
\endbibitem

\bibitem[\protect\citeauthoryear{Fatimah et~al.}{2022}]{9858160}
\begin{barticle}
\bauthor{\bsnm{Fatimah}, \binits{B.}},
\bauthor{\bsnm{Singh}, \binits{P.}},
\bauthor{\bsnm{Singhal}, \binits{A.}},
\bauthor{\bsnm{Pachori}, \binits{R.B.}}:
\batitle{Biometric identification from ecg signals using fourier decomposition and machine learning}.
\bjtitle{IEEE Transactions on Instrumentation and Measurement}
\bvolume{71},
\bfpage{1}--\blpage{9}
(\byear{2022})
\doiurl{10.1109/TIM.2022.3199260}
\end{barticle}
\endbibitem

\bibitem[\protect\citeauthoryear{Gupta and Mittal}{2020}]{gupta2020efficient}
\begin{barticle}
\bauthor{\bsnm{Gupta}, \binits{V.}},
\bauthor{\bsnm{Mittal}, \binits{M.}}:
\batitle{Efficient r-peak detection in electrocardiogram signal based on features extracted using hilbert transform and burg method}.
\bjtitle{Journal of the Institution of Engineers (India): Series B}
\bvolume{101}(\bissue{1}),
\bfpage{23}--\blpage{34}
(\byear{2020})
\end{barticle}
\endbibitem

\bibitem[\protect\citeauthoryear{Benitez et~al.}{2001}]{benitez2001use}
\begin{barticle}
\bauthor{\bsnm{Benitez}, \binits{D.}},
\bauthor{\bsnm{Gaydecki}, \binits{P.}},
\bauthor{\bsnm{Zaidi}, \binits{A.}},
\bauthor{\bsnm{Fitzpatrick}, \binits{A.}}:
\batitle{The use of the hilbert transform in ecg signal analysis}.
\bjtitle{Computers in biology and medicine}
\bvolume{31}(\bissue{5}),
\bfpage{399}--\blpage{406}
(\byear{2001})
\end{barticle}
\endbibitem

\bibitem[\protect\citeauthoryear{Kumar et~al.}{2018}]{kumar2018automated}
\begin{barticle}
\bauthor{\bsnm{Kumar}, \binits{M.}},
\bauthor{\bsnm{Pachori}, \binits{R.B.}},
\bauthor{\bsnm{Acharya}, \binits{U.R.}}:
\batitle{Automated diagnosis of atrial fibrillation ecg signals using entropy features extracted from flexible analytic wavelet transform}.
\bjtitle{Biocybernetics and Biomedical Engineering}
\bvolume{38}(\bissue{3}),
\bfpage{564}--\blpage{573}
(\byear{2018})
\end{barticle}
\endbibitem

\bibitem[\protect\citeauthoryear{Asgharzadeh-Bonab et~al.}{2020}]{asgharzadeh2020spectral}
\begin{barticle}
\bauthor{\bsnm{Asgharzadeh-Bonab}, \binits{A.}},
\bauthor{\bsnm{Amirani}, \binits{M.C.}},
\bauthor{\bsnm{Mehri}, \binits{A.}}:
\batitle{Spectral entropy and deep convolutional neural network for ecg beat classification}.
\bjtitle{Biocybernetics and Biomedical Engineering}
\bvolume{40}(\bissue{2}),
\bfpage{691}--\blpage{700}
(\byear{2020})
\end{barticle}
\endbibitem

\bibitem[\protect\citeauthoryear{{\`a}~Mougoufan et~al.}{2021}]{a2021three}
\begin{barticle}
\bauthor{\bsnm{Mougoufan}, \binits{J.B.B.}},
\bauthor{\bsnm{Fouda}, \binits{J.A.E.}},
\bauthor{\bsnm{Tchuente}, \binits{M.}},
\bauthor{\bsnm{Koepf}, \binits{W.}}:
\batitle{Three-class ecg beat classification by ordinal entropies}.
\bjtitle{Biomedical Signal Processing and Control}
\bvolume{67},
\bfpage{102506}
(\byear{2021})
\end{barticle}
\endbibitem

\bibitem[\protect\citeauthoryear{Frausto-Avila et~al.}{2025}]{frausto2025classification}
\begin{barticle}
\bauthor{\bsnm{Frausto-Avila}, \binits{M.}},
\bauthor{\bsnm{Ochoa-Elias}, \binits{M.}},
\bauthor{\bsnm{Manriquez-Amavizca}, \binits{J.P.}},
\bauthor{\bsnm{Carmen~Gonz{\'a}lez-L{\'o}pez}, \binits{M.}},
\bauthor{\bsnm{Ram{\'\i}rez-Garc{\'\i}a}, \binits{G.}},
\bauthor{\bsnm{Quiroz-Ju{\'a}rez}, \binits{M.A.}}:
\batitle{Classification of sers spectra for agrochemical detection using a neural network with engineered features}.
\bjtitle{Journal of Physics: Photonics}
\bvolume{7}(\bissue{2}),
\bfpage{025022}
(\byear{2025})
\end{barticle}
\endbibitem

\bibitem[\protect\citeauthoryear{Sahoo et~al.}{2017}]{sahoo2017multiresolution}
\begin{barticle}
\bauthor{\bsnm{Sahoo}, \binits{S.}},
\bauthor{\bsnm{Kanungo}, \binits{B.}},
\bauthor{\bsnm{Behera}, \binits{S.}},
\bauthor{\bsnm{Sabut}, \binits{S.}}:
\batitle{Multiresolution wavelet transform based feature extraction and ecg classification to detect cardiac abnormalities}.
\bjtitle{Measurement}
\bvolume{108},
\bfpage{55}--\blpage{66}
(\byear{2017})
\end{barticle}
\endbibitem

\bibitem[\protect\citeauthoryear{N et~al.}{2024}]{10563956}
\begin{bchapter}
\bauthor{\bsnm{N}, \binits{S.}},
\bauthor{\bsnm{S}, \binits{S.K.}},
\bauthor{\bsnm{Sharma~R}, \binits{R.}},
\bauthor{\bsnm{Sungheetha}, \binits{A.}},
\bauthor{\bsnm{R}, \binits{C.}},
\bauthor{\bsnm{Hamsanandhini}, \binits{S.}}:
\bctitle{Compact 1d self-operational neural networks with feature injection for global ecg classification}.
In: \bbtitle{2024 4th International Conference on Innovative Practices in Technology and Management (ICIPTM)},
pp. \bfpage{1}--\blpage{6}
(\byear{2024}).
\doiurl{10.1109/ICIPTM59628.2024.10563956}
\end{bchapter}
\endbibitem

\bibitem[\protect\citeauthoryear{Challagundla}{2024}]{challagundla2024advanced}
\begin{botherref}
\oauthor{\bsnm{Challagundla}, \binits{B.C.}}:
Advanced neural network architecture for enhanced multi-lead ecg arrhythmia detection through optimized feature extraction.
arXiv preprint arXiv:2404.15347
(2024)
\end{botherref}
\endbibitem

\bibitem[\protect\citeauthoryear{Acharya et~al.}{2017}]{acharya2017automated}
\begin{barticle}
\bauthor{\bsnm{Acharya}, \binits{U.R.}},
\bauthor{\bsnm{Fujita}, \binits{H.}},
\bauthor{\bsnm{Lih}, \binits{O.S.}},
\bauthor{\bsnm{Hagiwara}, \binits{Y.}},
\bauthor{\bsnm{Tan}, \binits{J.H.}},
\bauthor{\bsnm{Adam}, \binits{M.}}:
\batitle{Automated detection of arrhythmias using different intervals of tachycardia ecg segments with convolutional neural network}.
\bjtitle{Information sciences}
\bvolume{405},
\bfpage{81}--\blpage{90}
(\byear{2017})
\end{barticle}
\endbibitem

\bibitem[\protect\citeauthoryear{Acharya et~al.}{2016}]{acharya2016automated}
\begin{bchapter}
\bauthor{\bsnm{Acharya}, \binits{U.R.}},
\bauthor{\bsnm{Fujita}, \binits{H.}},
\bauthor{\bsnm{Adam}, \binits{M.}},
\bauthor{\bsnm{Lih}, \binits{O.S.}},
\bauthor{\bsnm{Hong}, \binits{T.J.}},
\bauthor{\bsnm{Sudarshan}, \binits{V.K.}},
\bauthor{\bsnm{Koh}, \binits{J.E.}}:
\bctitle{Automated characterization of arrhythmias using nonlinear features from tachycardia ecg beats}.
In: \bbtitle{2016 IEEE International Conference on Systems, Man, and Cybernetics (SMC)},
pp. \bfpage{000533}--\blpage{000538}
(\byear{2016}).
\bcomment{IEEE}
\end{bchapter}
\endbibitem

\end{thebibliography}


\end{document}